\documentclass[twoside]{elsart}
\usepackage{epsfig}
\usepackage{array}
\usepackage{amssymb,amsbsy,amsmath}
\newcommand{\skipit}[1]{}
\newcommand{\figref}[1]{fig.~\protect\ref{#1}}
\newcommand{\xref}[1]{\protect\ref{#1}}

\newcommand{\fmref}[1]{(\protect\ref{#1})}
\newlength{\CaptionWidth}
\newcommand{\mycaption}[2]{%
\begin{center}\begin{minipage}{\CaptionWidth}%
\caption[]{#1}\label{#2}%
\end{minipage}\end{center}%
}%
\def\leap{\raisebox{-.6ex}{$\stackrel {<}{\sim}$}} 
\def\V0{\stackrel{\circ}{V}}
\def\v0{\stackrel{\circ}{v}}
\def\half{{\frac{1}{2}}\;}

\newcommand{\MeV}{\mbox{\,MeV}}
\newcommand{\dint}{\mbox{d}}

\newcommand{\op}[1]{\ensuremath{%
    \fontdimen12\textfont3=2pt\fontdimen12\scriptfont3=1.4pt%
    \!\null\mathop{\vphantom{#1}\smash{#1}}\limits_{\sim}\null\!}}
\newcommand{\Operator}[1]{\smash{\raisebox{-1.1ex}{
$\!\!\stackrel{\displaystyle #1}{\sim}$}}}

\newcommand{\SmallMean}[1]{\langle\langle \; {#1}\; 
            \rangle\rangle}

\def\ket#1{\; | \, {#1} \, \rangle}

\newcommand{\pp}[2]{\frac{\partial \, {#1}}{\partial \, {#2}}\;}
\newcommand{\dd}[2]{\frac{{d}\, {#1}}{{d} {#2}}\;}

\begin{document}
%
\typeout{   --- >>>   ho paper   <<<   ---   }
\typeout{   --- >>>   ho paper   <<<   ---   }
\typeout{   --- >>>   ho paper   <<<   ---   }
%
%
\journal{PHYSICA A}
\begin{frontmatter}
\title{Investigations on finite ideal quantum gases}
 
\author{H.-J. Schmidt\thanksref{HJS}}
\author{ and J. Schnack\thanksref{JS}}
\address{Universit\"at Osnabr\"uck, Fachbereich Physik \\ 
         Barbarastr. 7, D-49069 Osnabr\"uck}

\thanks[HJS]{Email: hschmidt\char'100physik.uni-osnabrueck.de,\\
            WWW:~http://www.physik.uni-osnabrueck.de/makrosysteme/hjschmidt.htm}
\thanks[JS]{Email: juergen.schnack\char'100physik.uni-osnabrueck.de,\\
            WWW:~http://obelix.physik.uni-osnabrueck.de/$\sim$schnack}

\begin{abstract}
\noindent
Recursion formulae of the $N$-particle partition function,
the occupation numbers and its fluctuations are given using the
single-particle partition function. Exact results are presented
for fermions and bosons in a common one-dimensional harmonic
oscillator potential, for the three-dimensional harmonic
oscillator approximations are tested. Applications to excited
nuclei and Bose-Einstein condensation are discussed.

\vspace{1ex}


\vspace{1ex}

\end{abstract}
\end{frontmatter}
\raggedbottom
\section{Introduction and summary}

In thermodynamics ideal quantum gases are usually treated in the
grand canonical ensemble since all applications which were
important in the past, like the electron gas, phonons or
photons, deal with huge particle numbers, where the
thermodynamic limit is applicable.  Only the experimental attempts of
the last years to investigate finite Fermi and Bose systems and
to describe them in terms of thermodynamics called for
theoretical effort. Interesting finite Fermi systems are for
instance nuclei, which behave like a liquid drop and therefore
can undergo a first order phase transition \cite{Poc95}. On the
low excitation site of the caloric curve the nuclear systems
might be well described as an ideal Fermi gas in a common
harmonic oscillator potential. 

Small Bose systems became available through the development of
traps. Here the focus is on the Bose-Einstein condensation which
for instance could be found investigating alkali atoms in
magnetic traps \cite{AEM95,DMA95,BSH97}. Again the system can be
well described as an ideal quantum gas contained in an external
harmonic oscillator potential.

Theses finite quantum systems are characterized by a constant
particle number and may be represented by the canonical or
the micro-canonical ensemble depending on whether heat exchange with
the environment is possible or not.

In the following we present three formulae for the partition
function of the canonical ensemble, one being already known
\cite{BoF93}. All of them need the single-particle partition
function as input. In a one-dimensional harmonic oscillator the
result can be further simplified for fermions and bosons,
leading to the interesting result that both have the same
specific heat.

In practice the exact formulae for the partition functions can
be evaluated only for small particle numbers, say $N\leap 20$.
Therefore, approximations are derived in the third section,
which may be employed for larger $N$. They are tested against the
exact result for small $N$. In the last section we also discuss the
examples of small excited nuclei and Bose-Einstein condensation.

\section{Recursion formulae and generating functions}

Throughout this article we assume that the Hamilton operator is
a single particle operator, typically the kinetic energy and a
common potential or a mean field
\begin{eqnarray}
\label{E-2-1}
\op{H} = \sum_{n=1}^N \op{h}(n)
\ .
\end{eqnarray}
The partition function $Z_N$ for the canonical ensemble of $N$
particles can be recursively built starting with the
single-particle partition function \cite{BoF93}
\begin{eqnarray}
\label{E-2-2}
Z_N(\beta)
=
\frac{1}{N} \sum_{n=1}^N \; (\pm 1)^{n+1}\;
Z_1(n \beta) \; Z_{N-n}(\beta)
\ , \quad
Z_0(\beta) = 1
\ , \
\beta = \frac{1}{k_B T}
\ ,
\end{eqnarray}
where the upper sign in the sum stands for bosons, the lower sign
for fermions.

A second method uses a generating function to obtain
$Z_N$. Consider the grand canonical partition function
$Q(\beta,z)$
\begin{eqnarray}
\label{E-2-3}
Q(\beta,z)
=
\sum_{N=0}^\infty \;
z^N\; Z_N(\beta)
\ , \qquad
z = e^{\beta \mu}
\ ,
\end{eqnarray}
where $z$ is the fugacity and $\mu$ the chemical potential.
Multiplying $Q(\beta,z)$ with a second function $Y(\beta,z)$ 
\begin{eqnarray}
\label{E-2-4}
Y(\beta,z)
=
1 + \sum_{n=1}^\infty \;
(\pm 1)^{n}\;
z^n\; Z_1(n \beta)
\end{eqnarray}
and grouping the result with respect to powers of $z$ yields a
differential equation
\begin{eqnarray}
\label{E-2-5}
\frac{\left(1-Y(\beta,z)\right) \dint z}{z}
=
\frac{\dint Q(\beta,z)}{Q(\beta,z)}
\end{eqnarray}
which can be integrated obtaining
\begin{eqnarray}
\label{E-2-6}
\ln\left\{ Q(\beta,z) \right\} = \sum_{n=1}^{\infty}\;
(\pm 1)^{n+1}\;\frac{z^n}{n}\; Z_1(n \beta)
=: W(\beta,z)
\end{eqnarray}
or
\begin{eqnarray}
\label{E-2-7}
Q(\beta,z) = 
\exp\left\{ W(\beta,z) \right\}
\ .
\end{eqnarray}
Expanding this function, which expresses $Q(\beta,z)$ in terms
of $Z_1$, into powers of $z$ gives the $Z_N(\beta)$ as coefficients.
This derivation is equivalent to first differentiating $Q(\beta,z)$
eq. \fmref{E-2-3} with respect to $z$, inserting
\fmref{E-2-2} and then regrouping the sum \cite{Bae98}.

There is even a third method which represents $Z_N(\beta)$
explicitly in terms of $Z_1$
\begin{eqnarray}
\label{E-2-8}
Z_N(\beta)
=
\frac{1}{N!} 
\sum_{\nu=0}^{2^{N-1}-1} \; 
(\pm 1)^{N-k[\nu]}\;
N[\nu]
\prod_{n\in K[\nu]}\;
Z_1(n \beta)
\ .
\end{eqnarray}
We explain this method with the help of $Z_3(\beta)$ and table
\xref{T-1-1}. 
\begin{table}[ht!]
\begin{center}
\begin{tabular}{c|c|c|c|c||c}
$\nu$ & $[\nu]$ & $K[\nu]$ & $k[\nu]$ & $N[\nu]$ & result \\
\hline\hline
0 & [000]   & (1,1,1)& 3 &$ 1^0 2^0 3^0 = 1$ & 
$(\pm 1)^0 1 Z_1^3(\beta)$\\
1 & [100]   & (2,1)  & 2 &$ 1^1 2^0 3^0 = 1$ & 
$(\pm 1)^1 1 Z_1(2 \beta) Z_1(\beta)$\\
2 & [010]   & (1,2)  & 2 &$ 1^0 2^1 3^0 = 2$ & 
$(\pm 1)^1 2 Z_1(\beta) Z_1(2 \beta)$\\
3 & [110]   & (3)    & 1 &$ 1^1 2^1 3^0 = 2$ & 
$(\pm 1)^2 2 Z_1(3 \beta)$\\
\hline
\end{tabular}
\vspace*{5mm}
\end{center}
\mycaption{Example of the explicit method eq. \fmref{E-2-8} for
$N=3$.}{T-1-1}
\end{table}
The method starts with the dual representation of $\nu$ running
from zero to $2^{N-1}-1$. This string $[\nu]$ is filled with
zeros at the end to get as many digits as particles, in our
example three. Then this string is divided into $k[\nu]$ substrings
each substring closing with exactly one zero, in our example for $\nu=1$
the string $100$ is grouped into $10$ and $0$. $K[\nu]$ is the
$k[\nu]$-tuple which consists of the lengths of those
substrings. The last factor involved is $N[\nu]$ which is the
product of all integers from $1$ to $N$ taken to the power given
by the respective digit in the string $[\nu]$. Adding up the
results of our example leads to
\begin{eqnarray}
\label{E-2-9}
Z_3(\beta)
=
\frac{1}{3!} 
\left(
Z_1^3(\beta) \;
\pm 3 Z_1(\beta) Z_1(2 \beta) \;
+ 2 Z_1(3 \beta)
\right)
\ .
\end{eqnarray}
The proof of \fmref{E-2-8} is accomplished by regrouping the
contributions $\nu$ with respect to $[\nu]$, where the first
group contains all terms where $[\nu]$ ends with $00$, the
second where $[\nu]$ ends with $010$, the third where $[\nu]$
ends with $0110$ and so on. These groups can be identified with
the summands of the recursion formula \fmref{E-2-2}. The
coefficients of the product $\prod Z_1(n \beta)$ in
\fmref{E-2-8} satisfy a number of combinatorial relations which
will be investigated elsewhere. Here we only note, that the sum
of all coefficients amounts to $0$ in the fermionic case and to
$1$ in the bosonic case.

Following the very same idea that led to \fmref{E-2-2} one can
derive recursion formulae for the occupation numbers of
single--particle energy eigenstates $\ket{n}$ with energy
$\varepsilon_n$
\cite{Bor95}
\begin{eqnarray}
\label{E-2-14}
N_n
:=
\SmallMean{\Operator{a}_n^+\Operator{a}_n}
=
\frac{1}{Z_N(\beta)}
\sum_{m=1}^N \; (\pm 1)^{m+1}\;
\exp\left\{ -m\;\beta\varepsilon_n\right\} \;
Z_{N-m}(\beta)
\ .
\end{eqnarray}
Any single--particle density $\rho_N(\beta)$ (e.g. the spatial
density) is given by
\begin{eqnarray}
\label{E-2-15}
\rho_N(\beta)
=
\frac{1}{Z_N(\beta)}
\sum_{n=1}^N \; (\pm 1)^{n+1}\;
\rho_1(n \beta)\;
Z_1(n \beta) \; Z_{N-n}(\beta)
\ ,
\end{eqnarray}
which follows from equation \fmref{E-2-14} using the density $\rho_1$ in
one-particle space. For the fermion case the fluctuations are
well known to be
\begin{eqnarray}
\label{E-2-16}
\delta N_n
:=
\SmallMean{\left(\Operator{a}_n^+\Operator{a}_n\right)^2}
-\SmallMean{\Operator{a}_n^+\Operator{a}_n}^2
=
\SmallMean{\Operator{a}_n^+\Operator{a}_n}
\left(1 - \SmallMean{\Operator{a}_n^+\Operator{a}_n}\right)
\end{eqnarray}
for bosons the fluctuations are given with the help of
\begin{eqnarray}
\label{E-2-17}
\SmallMean{\left(\Operator{a}_n^+\Operator{a}_n\right)^2}
&=&
\frac{1}{Z_N(\beta)}
\sum_{m=2}^N \; (m-1)\;
\exp\left\{ -m\;\beta\varepsilon_n\right\} \;
Z_{N-m}(\beta)
\ .
\end{eqnarray}

\subsection{Exact solution for the one-dimensional harmonic oscillator}

In the case of a one-dimensional harmonic oscillator potential
the $N$-particle partition functions can be given explicitly,
they are (see also \cite{AuK46} for bosons)
\begin{eqnarray}
\label{E-2-10}
Z_N(\beta)
&=&
\exp\left(-\beta E_0(N)\right)
\prod_{n=1}^N\;
\frac{1}{1 - \exp(-n\; \beta \hbar\omega)} 
\ ,
\end{eqnarray}
which differ for bosons and fermions only in $E_0$
\begin{eqnarray}
\label{E-2-11}
E_0(N) &=& N \frac{\hbar\omega}{2}
\qquad\mbox{for bosons}
\\
E_0(N) &=& N^2 \frac{\hbar\omega}{2}
\qquad\mbox{for fermions}
\ .
\end{eqnarray}
The mean energy turns out as
\begin{eqnarray}
\label{E-2-12}
E_N(T) = 
E_0(N) +
\sum_{n=1}^N\;
n \frac{\hbar\omega}{2}\;
\left[
\mbox{coth}\left( n \frac{\beta\hbar\omega}{2} \right)
- 1 \right]
\ .
\end{eqnarray}
It is interesting to realize that ideal Fermi and Bose gases
contained in a one-dimensional harmonic oscillator have the same
specific heat
\begin{eqnarray}
\label{E-2-13}
c =
\frac{1}{N}
\left(
\pp{E(T)}{T}
\right)_{\omega}
=
\frac{k_B}{N\; T^2}
\sum_{n=1}^N\;
\left(n \frac{\hbar\omega}{2}\right)^2
\frac{1}{\mbox{sinh}^2\left( n \frac{\beta\hbar\omega}{2} \right)}
\ ,
\end{eqnarray}
which is due to the equidistant spacing of energy levels and is,
for instance, not valid in the case of a one-dimensional box.

\subsection{Exact solution for the one-dimensional box}

Since the formulae \fmref{E-2-2}, \fmref{E-2-6} and
\fmref{E-2-8} for the $N$-particle partition function rely on
the knowledge of the single-particle 
partition function it is always valuable to have $Z_1$ 
explicitly. Considering a particle with mass $m$ in a
one-dimensional box of length $L$ the energy eigenvalues are
\begin{eqnarray}
\label{E-2-20}
E_n = \frac{\hbar^2}{2 m}\;\left(\frac{\pi\;n}{L}\right)^2
\qquad \mbox{and}\quad
x:=\beta\;
\frac{\hbar^2}{2 m}\;\left(\frac{\pi}{L}\right)^2
\end{eqnarray}
will denote the dimensionless inverse temperature.
Then the partition function for the one-dimensional box can be
calculated as
\begin{eqnarray}
\label{E-2-22}
Z_1(\beta)
&=&
\sum_{n=1}^\infty\;
\mbox{e}^{-x\; n^2}
=
\half \left(
\vartheta_3(0,\mbox{e}^{-x}) -1
\right)
=
\sqrt{\frac{K(m)}{2\pi}} - \half
\ .
\end{eqnarray}
In three dimensions it is just the third power.
$\vartheta_3$ is the third elliptic $\vartheta$-function
and $K$ the complete elliptic integral \cite{Rai60}
\begin{eqnarray}
\label{E-2-23}
\vartheta_3(u,q)
&=&
1 + 2 \sum_{n=0}^\infty\;q^{n^2}\;
\cos(2 n u)
\ ,\\
K(m)
&=&
\int_{0}^{\pi/2} \dint\theta\;
\left(1 - m \; \sin^2(\theta)\right)^{-\half}
\nonumber
\ .
\end{eqnarray}
The variable $m$ is related to the dimensionless inverse
temperature $x$ by
\begin{eqnarray}
\label{E-2-25}
m
=
\left(
\frac{\vartheta_2(0,\mbox{e}^{-x})}{\vartheta_3(0,\mbox{e}^{-x})}
\right)^4
\ ,\quad
\vartheta_2(u,q)
=
2 q^{\frac{1}{4}} \;
\sum_{n=0}^\infty\;q^{n(n+1)}\;
\cos(2 (n+1) u)
\ .
\end{eqnarray}
In order to calculate the mean energy the first derivative of
$Z_1$ is needed which reads
\begin{eqnarray}
\label{E-2-26}
\dd{Z_1}{x}
=
- \frac{\pi^{-\frac{5}{2}}}{\sqrt{2}}\;
K^{\frac{3}{2}}(m)\;
\left[
E(m) - (1-m) K(m)
\right]
\ ,
\end{eqnarray}
with $E$ the complete elliptic integral
\begin{eqnarray}
\label{E-2-27}
E(m)
=
\int_{0}^{\pi/2} \dint\theta\;
\left(1 - m \; \sin^2(\theta)\right)^{\half}
\ .
\end{eqnarray}

\section{Approximations}

\subsection{Fermions}

The approximation rests on the fact that a three dimensional
harmonic oscillator can be viewed as three independent one
dimensional oscillators, one for each spatial coordinate. Then
the idea is to apply the Pauli principle for excited states only
to one of the oscillators, i.e. directions in space. The
approximation of the partition function becomes \cite{Sch96}
\begin{eqnarray}
\label{E-3-1}
\tilde{Z}_N(\beta)
=
\exp\left(-\beta E_0(N)\right)\;
&&
\left(\frac{1}{1 - \exp(-\beta \hbar\omega)}\right)^{2N}\;
\\
&\times&
\prod_{n=1}^N\;
\frac{1}{1 - \exp(-n\; \beta \hbar\omega)} 
\nonumber
\end{eqnarray}
with $E_0(N)$ being the correct ground state energy. The
approximate mean energy
\begin{eqnarray}
\label{E-3-2}
\tilde{E}_N(T) = 
E_0(N) 
&+&
2 N \frac{\hbar\omega}{2} \left[
\mbox{coth}\left( \frac{\beta\hbar\omega}{2} \right)
- 1 \right]
\\
&+&
\sum_{n=1}^N\;
n \frac{\hbar\omega}{2}\;
\left[
\mbox{coth}\left( n \frac{\beta\hbar\omega}{2} \right)
- 1 \right]
\nonumber
\end{eqnarray}
has besides the ground state energy three parts, two for the two
directions where the Pauli principle is not applied and one
where it is applied. Therefore the specific heat is a sum of two
contributions for distinguishable particles in a common
one-dimensional oscillator field and one for fermions in a
common one-dimensional oscillator field.  Although the partition
function is not complete in this approximation the resulting
mean energy turns out to deviate from the correct result by less
then 1\%, see example \figref{F-3-1}.  Only for $N=2$ and $N=3$ the
deviation is larger. The high temperature limit is correct.

\begin{figure}[ht!]
\unitlength1mm
\begin{center}
\begin{picture}(120,45)
\put(0,0){\epsfig{file=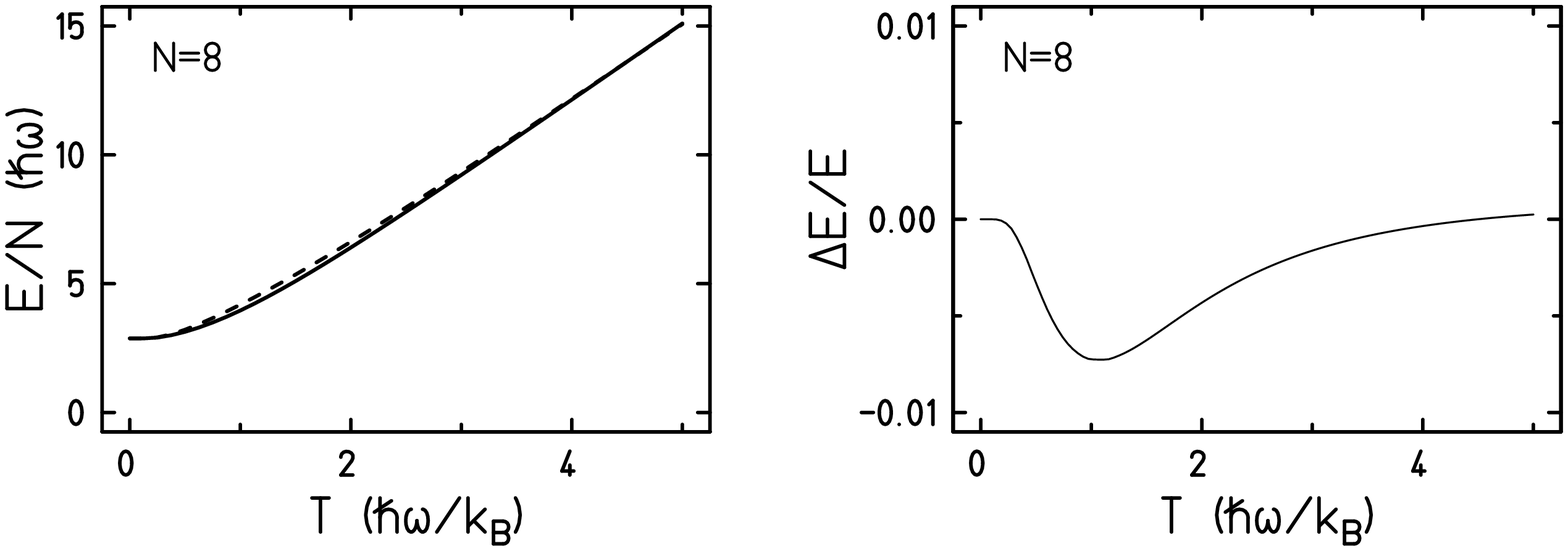,width=120mm}}
\end{picture}
\end{center}
\mycaption{L.h.s: mean energy of fermions in a three-dimensional
oscillator, the solid line displays the exact result, the dashed
line the approximation \fmref{E-3-2}. R.h.s.: relative deviation
between approximation and exact result.}{F-3-1}
\end{figure} 

\subsection{Bosons}

Unfortunately an approximation like \fmref{E-3-1} cannot be
derived for bosons, here we try to derive an appropriate power
series. For the ideal Bose gas in a three-dimensional oscillator
potential we have 
\begin{eqnarray}
\label{E-3-3}
Z_1
=
\frac{x^3}{(1-x^2)^3}
\ ,\qquad\mbox{with}\ 
x:=\mbox{e}^{-\beta\hbar\omega/2}
\ .
\end{eqnarray}
Obviously, $Z_N$ will be a rational function of $x$. A closer
inspection shows that it has the structure
\begin{eqnarray}
\label{E-3-4}
Z_N(x)
=
\frac{x^{3N}}{\prod_{n=1}^{N}(1-x^{2n})^3}\;
P_N(x)
\ ,\qquad\mbox{where}\ 
P_N(x)
=
\sum_{m=0}^{L_N} p_m\; x^{2m}
\end{eqnarray}
is an even polynomial with non-negative coefficients where for
$N\ge 4$ the degree is $2 L_N=(3 N^2 - 7 N+10)$. The first
$(N+1)$ coefficients $p_0, p_1, \dots, p_N$ are independent of
$N$, for example,
\begin{eqnarray}
\label{E-3-5}
P_N(x)
=
1 + 3 x^4 + \cdots
\qquad\mbox{for}\ N > 1
\ .
\end{eqnarray}
Moreover, the following identities hold (the primes denote
derivatives): 
\begin{eqnarray}
\label{E-3-6}
P_N(1) &=& (N!)^2
\\
P_N^{\prime}(1) &=& \frac{3}{2} N (N-1) (N!)^2
\nonumber \\
P_N^{\prime\prime}(1) &=& \frac{1}{12} (27 N^2 - 23 N - 8) 
N (N-1) (N!)^2
\nonumber \\
P_N^{\prime\prime\prime}(1) &=& \frac{1}{8} (27 N^4 - 42 N^3 - 9
N^2 +16 N -20) N (N-1) (N!)^2
\nonumber
\ .
\end{eqnarray}
These identities may be proved by deriving an analogous
recursion relation for the $P_N$ from eq. \fmref{E-2-2}.

In order to approximate $Z_N$ we consider the function
\begin{eqnarray}
\label{E-3-7}
f_N(x)
=
\ln\left[ \ln \left( P_N(x) \right) \right]
\ ,
\end{eqnarray}
which is analytic in a neighborhood of the real positive axis and
has a logarithmic singularity at $x=0$ due to
\fmref{E-3-5}. Since $f_N(x)-4\ln(x)$ behaves rather smoothly in
the physically relevant interval $x\in(0,1]$ it appears sensible
to use a Taylor series approximation
\begin{eqnarray}
f_N(x)
&=:&
4 \ln(x) + g_N(x)
\nonumber
\\
&\approx&
4 \ln(x) + 
\sum_{\nu=0}^K\;\frac{g_{\nu}}{\nu !}
(x-1)^{\nu}
\label{E-3-8}
\ .
\end{eqnarray}
The values of the $\nu$-th derivatives $f_N^{(\nu)}(1),
\nu=0,\dots,K,$ and hence of $g_{\nu}$, can be expressed in a
straight forward way as functions of $P_N^{(\nu)}(1)$ and, using
identities of the form
\fmref{E-3-6}, finally as known functions of $N$, albeit of increasing
complexity. We choose $K=3$ and $K=4$ and obtain approximations
$\tilde{Z}_N$, $\tilde{E}_N$ and $\tilde{c}_N$, where the latter
is compared to exact results in figure \xref{F-3-2}.

\begin{figure}[ht!]
\unitlength1mm
\begin{center}
\begin{picture}(120,45)
\put(0,0){\epsfig{file=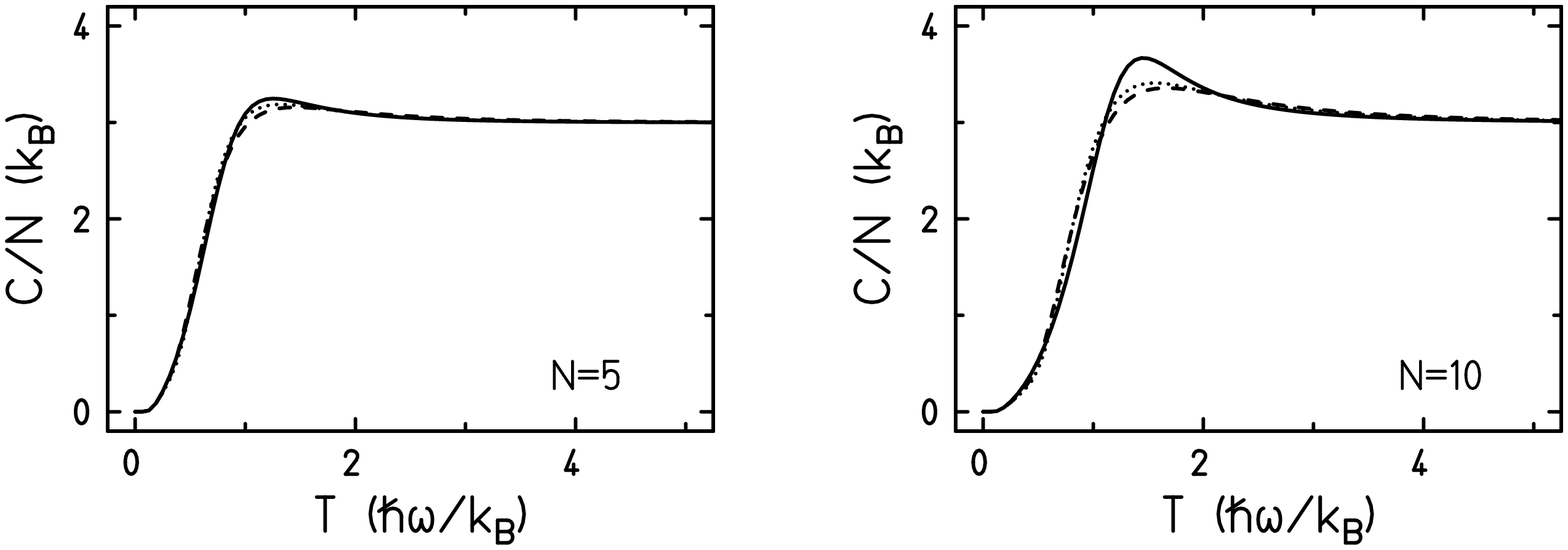,width=120mm}}
\end{picture}
\end{center}
\mycaption{Specific heat of bosons in a three-dimensional
oscillator, the solid line displays the exact result, the dashed
line the approximation \fmref{E-3-8} with $K=3$, the dotted with
$K=4$.}{F-3-2}
\end{figure} 

One sees that the approximation already for small $K$ reflects
the exact behaviour qualitatively correct. In the high and low
temperature regime it works also quantitatively well, but the
maximum is described only roughly both in position and heights.
For larger particle numbers more and more terms of the Taylor
expansion \fmref{E-3-8} would have to be taken into account.

\section{Applications}

\subsection{Fermions}

\begin{figure}[ht!]
\unitlength1mm
\begin{center}
\begin{picture}(120,60)
\put( 0,0){\epsfig{file=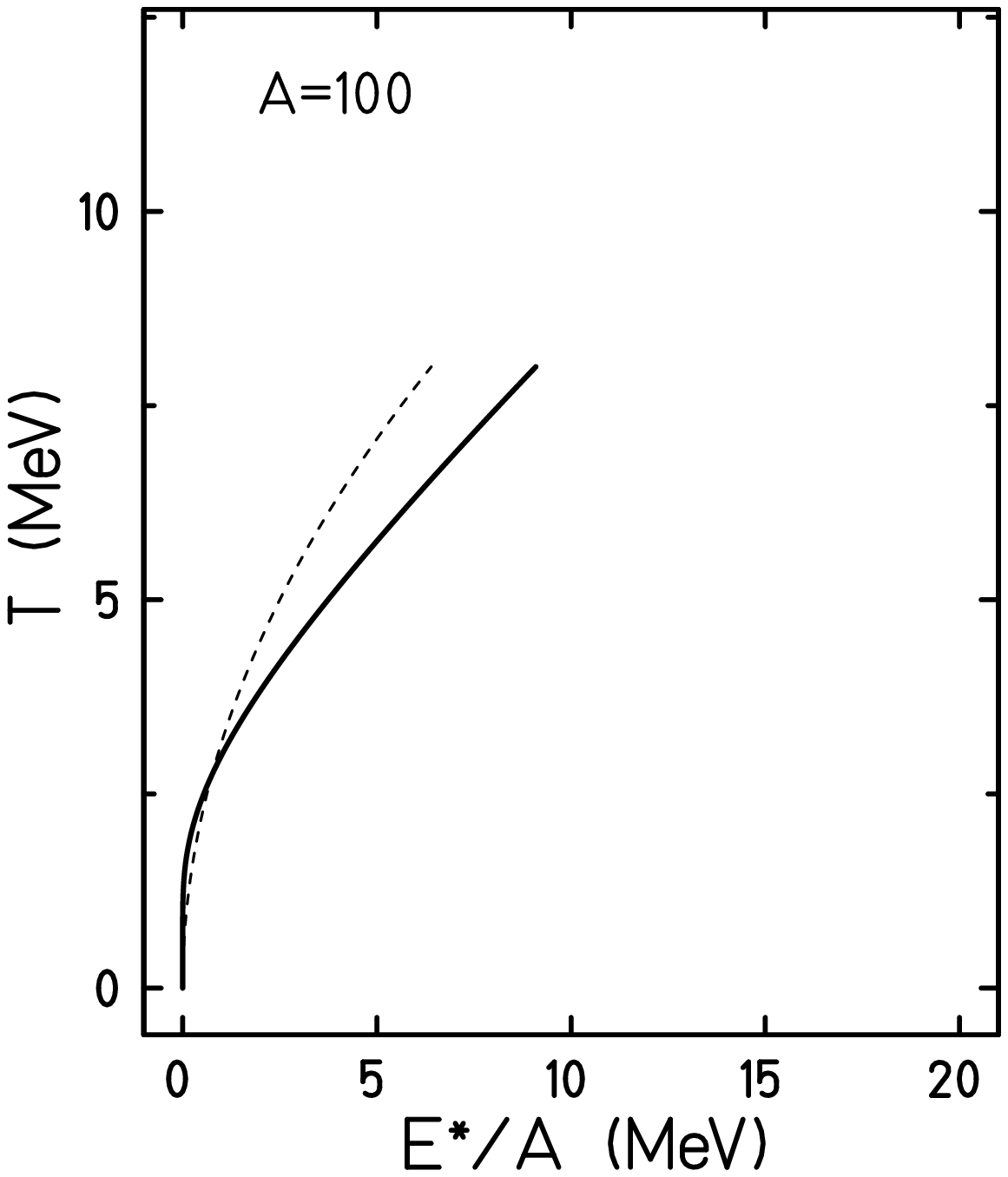,width=52mm}}
\put(65,0){\epsfig{file=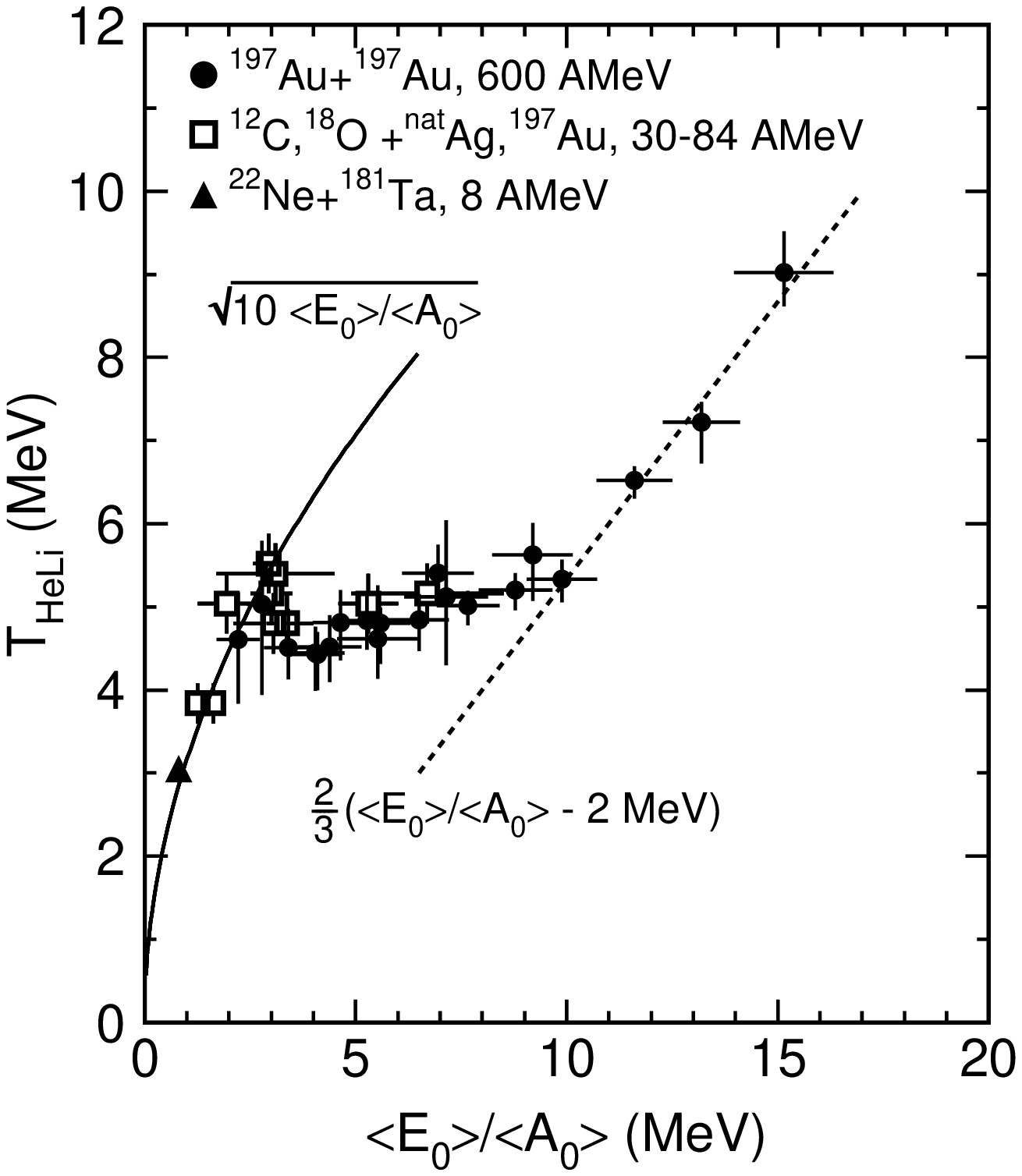,width=55mm}}
\end{picture}
\end{center}
\mycaption{L.h.s.: relation between excitation energy and
temperature for the harmonic oscillator (solid line) and the
free Fermi gas (dashed line). R.h.s.: experimentally determined
relation, figure taken from \cite{Poc95}.}{F-4-3}
\end{figure} 

Approximation \fmref{E-3-2} enables us to evaluate the low
temperature behavior of the excitation energy of nuclei
consisting of for instance $A=100$ nucleons and compare it to the
approximation of the free Fermi gas commonly employed in nuclear
physics. Figure \xref{F-4-3} shows on the l.h.s. as a solid line
the result obtained for the common harmonic oscillator with a
frequency of
\begin{eqnarray}
\label{E-4-1}
\hbar \omega
\approx
41\MeV\; A^{-\frac{1}{3}}
\ ,
\end{eqnarray}
which is obtained if one demands a correct root mean square
radius \cite{RiS80}. The dashed line displays the respective
dependence for the free Fermi gas with an inverse level density
parameter of $10\MeV$. One clearly sees that for very small
excitations the temperature rises much faster with excitation
energy if calculated in the oscillator shell model. Then the
curve bends earlier towards the high temperature limit $c=3k_B$,
meaning that the temperature now grows less with excitation
energy. In practice this observation is still beyond experimental
verification as can be anticipated from the right hand side,
where an attempt to determine the complete caloric curve of the
nuclear liquid-gas phase-transition is displayed \cite{Poc95}.

\subsection{Bosons}

The observation of Bose-Einstein condensation of very cold gases
contained in magnetic traps has stimulated much theoretical work
in this field during the last years (see for instance
\cite{KeD96,GrH96,KiT96,WiW97} and references therein). Since
the investigated gases are very dilute and the trap can be
modeled by a harmonic oscillator potential these systems can be
viewed as ideal Bose gases in common harmonic oscillator fields.

Before going into details let us first remind some known results
\cite{Hua87}. Rigorously speaking a phase transition towards the
Bose-Einstein condensate does not happen in an external
potential since the chemical potential is always non-zero for
finite temperatures. Only for a vanishing potential (infinite
volume limit) it is zero for all temperatures below $T_c$. It is
also true that the phase transition does only occur for
dimensions $d > 2$. 

\begin{figure}[ht!]
\unitlength1mm
\begin{center}
\begin{picture}(120,85)
\put(0,0){\epsfig{file=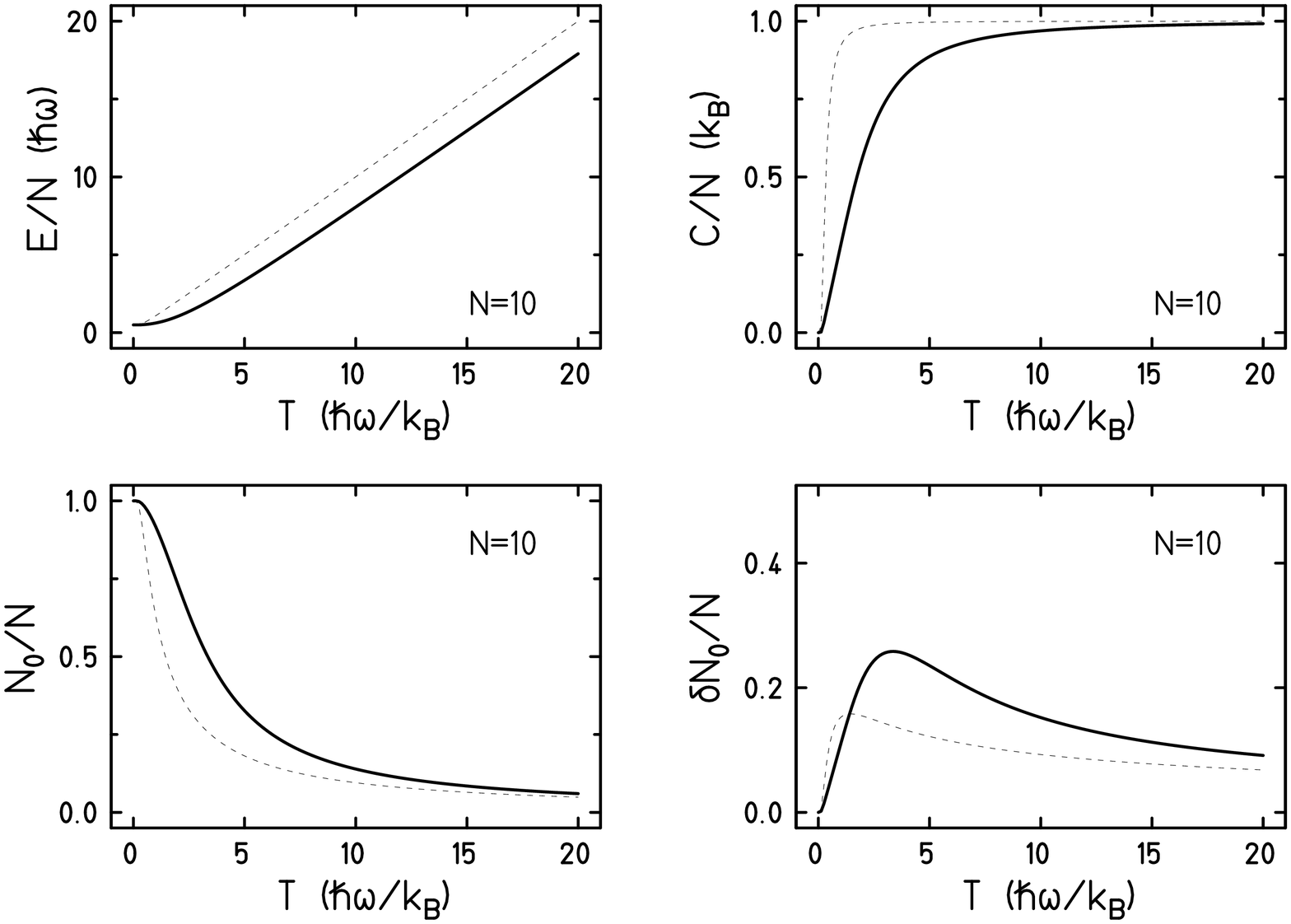,width=120mm}}
\end{picture}
\end{center}
\mycaption{One-dimensional harmonic oscillator: mean energy,
specific heat, ground state occupation number and its
fluctuation for bosons (solid line) and distinguishable
particles (dashed line).}{F-4-1}
\end{figure} 

\begin{figure}[ht!]
\unitlength1mm
\begin{center}
\begin{picture}(120,90)
\put(0,0){\epsfig{file=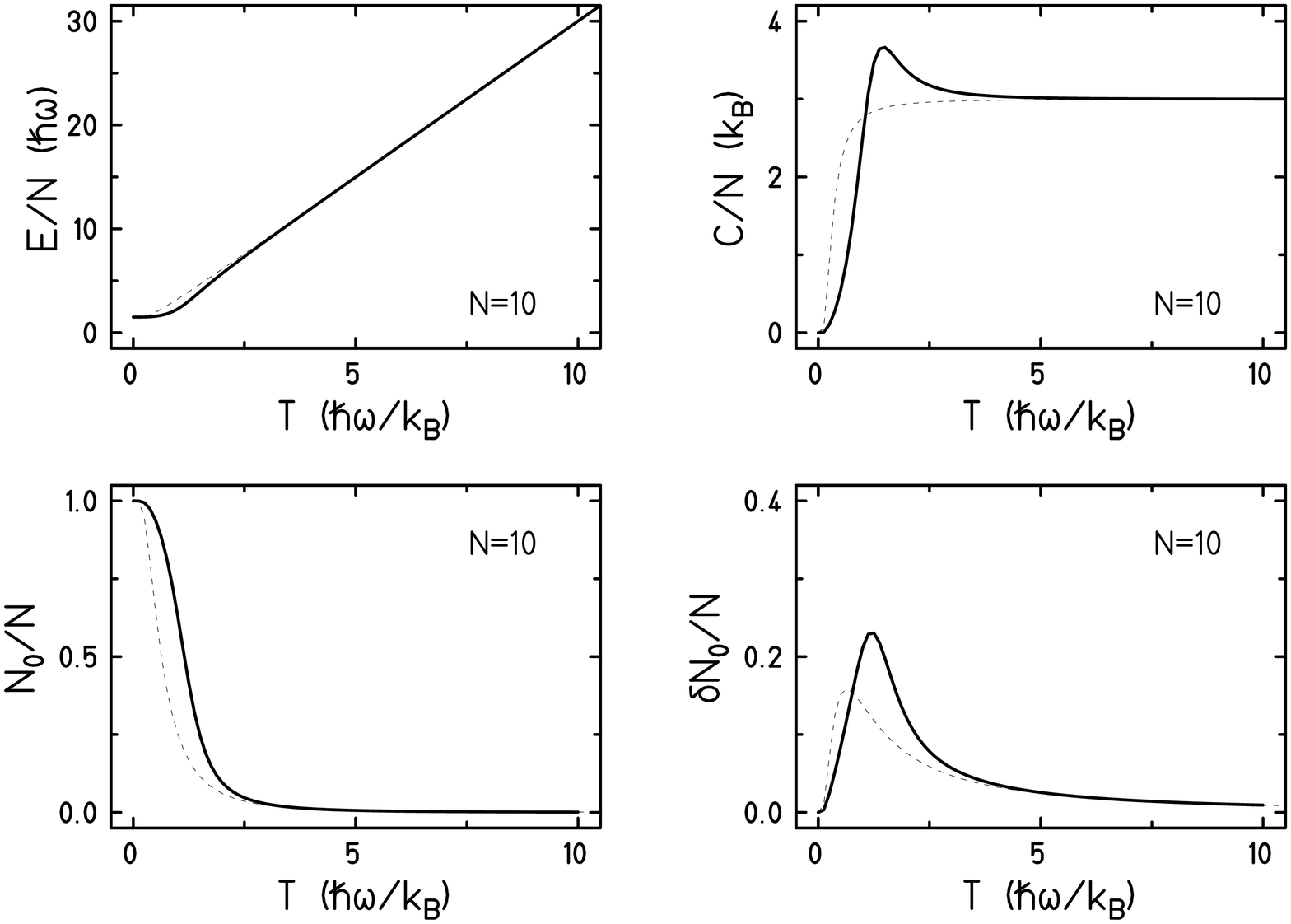,width=120mm}}
\end{picture}
\end{center}
\mycaption{Three-dimensional harmonic oscillator: mean energy,
specific heat, ground state occupation number and its
fluctuation for bosons (solid line) and distinguishable
particles (dashed line).}{F-4-4}
\end{figure} 

Dealing with finite systems one has to relax the rigorous point
of view and look for criteria that one would like to identify a
"smooth Bose-Einstein condensation". A first natural criterion
is the maximum in the specific heat \cite{PaP77,KiT96}, which is
a kink in the free three-dimensional case because the mean
energy jumps at $T_c$. Looking at figures \xref{F-4-1} and
\xref{F-4-4} (upper right corner) one sees that the specific heat
exhibits a clear maximum in the three-dimensional oscillator
whereas it does not in the one-dimensional oscillator. Using the
specific heat as a criterion for Bose-Einstein condensation the
property that it does only occur for dimensions larger than two
is maintained.

A second possible criterion would be the ground state occupation
number $N_0$ and its fluctuation $\delta N_0$. One could argue
that the ground state should be macroscopically occupied below a
certain temperature $T_c$ \cite{KeD96}, but as figures
\xref{F-4-1} and \xref{F-4-4} (lower left) show, that happens
also with distinguishable particles (dashed line). The same is
true if one would take the existence of a maximum in the
fluctuations of the ground state occupation number as a
criterion. Also here the mere existence happens for
distinguishable particles too, see figures \xref{F-4-1} and
\xref{F-4-4} (lower right). One also realizes that in this
respect there is no difference between the one- or
three-dimensional case.

\begin{figure}[ht!]
\unitlength1mm
\begin{center}
\begin{picture}(120,100)
\put(0,50){\epsfig{file=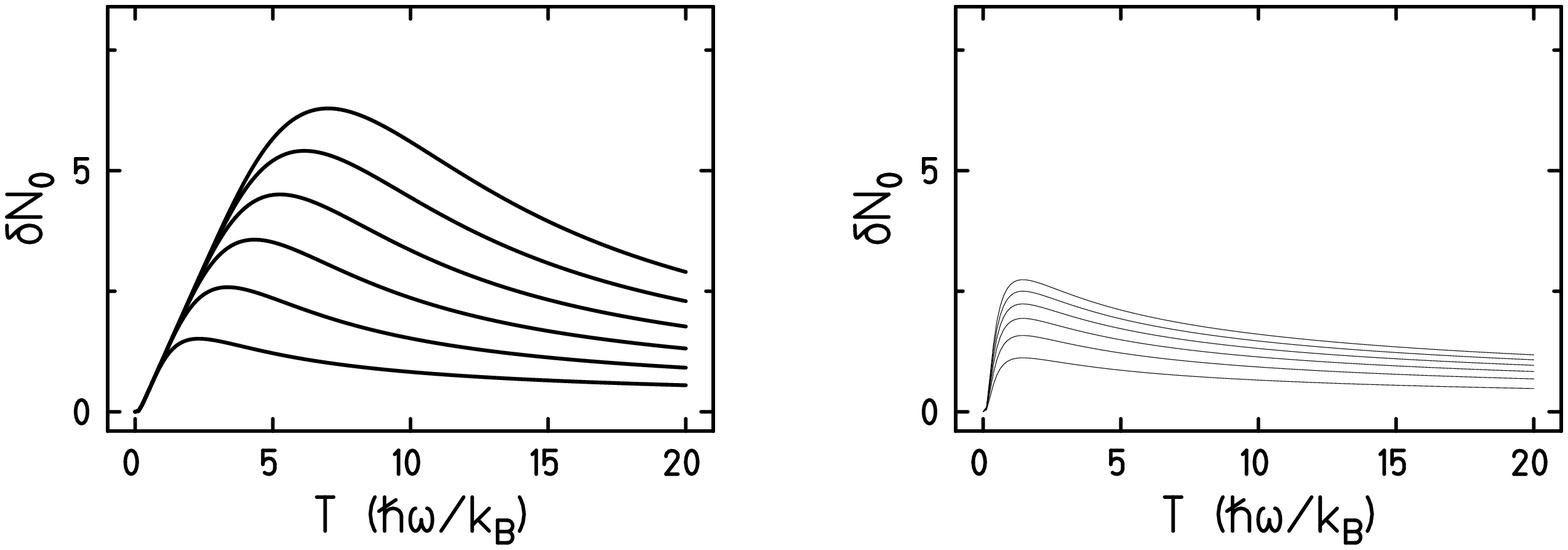,width=120mm}}
\put(0, 0){\epsfig{file=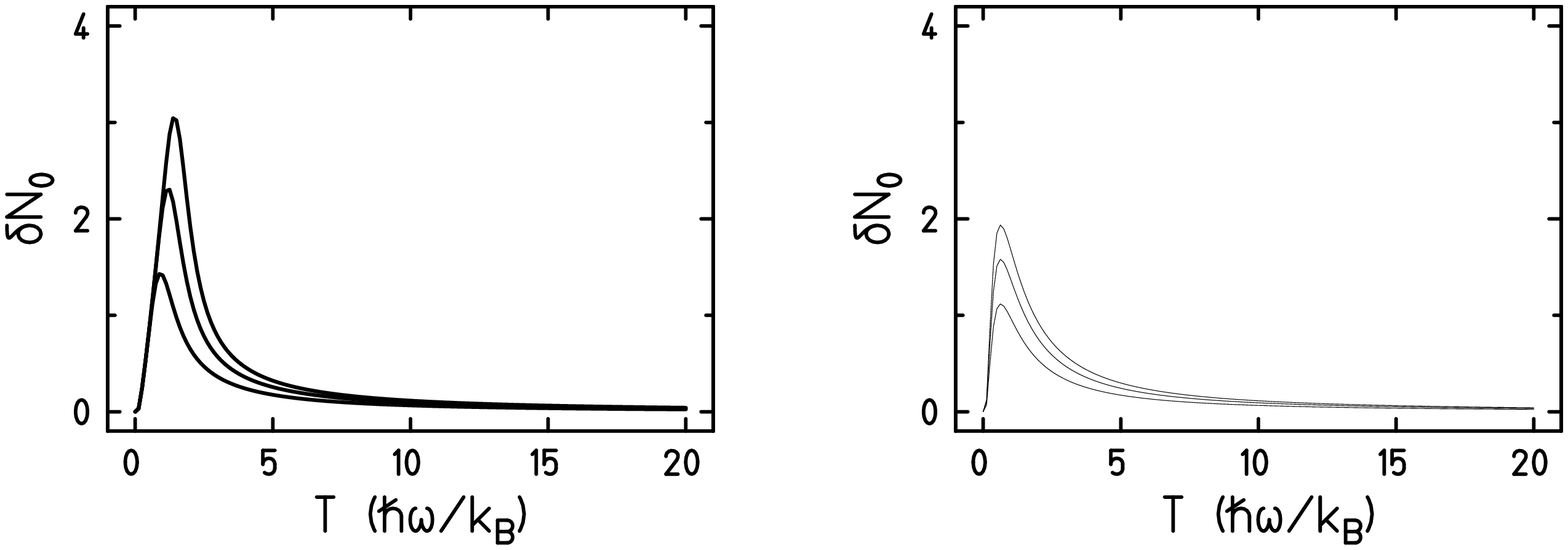,width=120mm}}
\end{picture}
\end{center}
\mycaption{Fluctuations of the ground state occupation number in
a one-dimensional (upper) and a three-dimensional harmonic
oscillator (lower); lines from bottom to
top starting with $N=5$ in steps of $5$. L.h.s. bosons,
r.h.s. distinguishable particles.}{F-4-2}
\end{figure} 

The possible key point in order to distinguish between bosons
and distinguishable particles is the dependence of the magnitude
of the fluctuations on the particle number. For bosons in a
one-dimensional harmonic oscillator the fluctuations $\delta
N_0$ scale with ${N}$ (see also \cite{GrH96}), whereas for
distinguishable particles they scale with $\sqrt{N}$. This can
also be seen in \figref{F-4-2} (upper figures).

In an experimental situation one of course knows whether one is
dealing with bosons or not. The experimental interest then is to
bring as many bosons as possible into the ground state. Comparing
the one-dimensional (\figref{F-4-2} upper figures) and
three-dimensional case (\figref{F-4-2} lower figures) one
realizes that for the same particle number this happens at
higher temperatures in one dimension.

{\bf Acknowledgments}\\[5mm]
J.S. would like to thank J\"orn Knoll and Rolf Fauser for
helpful advise. We thank Klaus B\"arwinkel for carefully reading
the manuscript.


\begin{thebibliography}{99}
\bibitem{Poc95}
	J.~Pochodzalla et al.,
	Phys. Rev. Lett. {\bf 75} (1995) 1040
%
%
\bibitem{AEM95}
	M.H. Anderson, J.R.~Ensher, M.R.~Matthews, C.E.~Wieman,
	E.A. Cornell,
	Science {\bf 269} (1995) 198
%
%
\bibitem{DMA95}
	K.B. Davis, M.-O. Mewes, M.R.~Andrews, N.J.~van Druten,
	D.S.~Durfee, D.M.~Kurn, W.~Ketterle,
	Phys. Rev. Lett. {\bf 75} (1995) 3969
%
%
\bibitem{BSH97}
	C.C.~Bradley, C.A.~Sackett, R.G.~Hulet,
	Phys. Rev. Lett. {\bf 78} (1997) 985
%
%
\bibitem{BoF93}
	P. Borrmann, G. Franke,
	J. Chem. Phys. {\bf 98} (1993) 2484
%
%
\bibitem{Bae98}
	K. B\"arwinkel, private communication
%
%
%
\bibitem{Bor95}
	P. Borrmann,
	dissertation, University of Oldenburg (1995)
	and preprint cond-mat/9412117
%
%
\bibitem{AuK46}
	F.C. Auluck, D.S. Kothari,
	Proc. Cambridge Philos. Soc. {\bf 42} (1946) 272
%
%
\bibitem{Rai60}
	E.D. Rainville,
	{\it Special functions, }
	Chelsea Publishing Company, New York (1960)
%
%
\bibitem{Sch96}
	J. Schnack, dissertation, TH Darmstadt (1996)
%
%
\bibitem{RiS80}
	P. Ring, P. Schuck,
	{\it The Nuclear Many-Body Problem,}
	Texts and monographs in physics,
	Springer (1980)
%
%
%
\bibitem{KeD96}
	W. Ketterle, N.J. van Druten,
	Phys. Rev. {\bf A54} (1996) 656
%
%
\bibitem{GrH96}
	S. Grossmann, M. Holthaus,
	Phys. Rev. {\bf E54} (1996) 3495;
	Phys. Rev. Lett. {\bf 79} (1997) 3557;
	Optics Express {\bf 1} (1997) 262
%
%
\bibitem{KiT96}
	K.Kirsten, D.J. Toms,
	Phys. Rev. {\bf A54} (1996) 4188
%
%
\bibitem{WiW97}
	M. Wilkens, C. Weiss,
	Journal of Modern Optics {\bf 44} (1997) 1801;
	Optics Express {\bf 1} (1997) 272
%
%
\bibitem{Hua87}
	K. Huang,
	{\it Statistical mechanics, }
	John Wiley \& Sons, New York (1987)
%
%
\bibitem{PaP77}
	H.R. Pajkowski, R.K. Pathria,
	J. Phys. {\bf A10} (1977) 561
%
%
%
%
%
%
%
%
%
%
%
\end{thebibliography}
\end{document}